# Scalable near-infrared graphene plasmonic resonators exhibiting strong non-local and electron quantization effects

Joel. F. Siegel[1τ], Jonathan H. Dwyer[2τ], Anjali Suresh[3], Nathaniel S. Safron[3], Margaret Fortman[1], Chenghao Wan[3,4], Jonathan W. Choi[3], Wei Wei[3], Vivek Saraswat[3], Wyatt A. Behn[1], Mikhail A. Kats[1,3,4], Michael S. Arnold[3], Padma Gopalan[2,3], and Victor W. Brar[1]*

*Address correspondence to: vbrar@wisc.edu
τAuthors contributed equally

[1]Department of Physics, University of Wisconsin – Madison, 1150 University Ave., Madison, WI 53706
[2]Department of Chemical and Biological Engineering, University of Wisconsin – Madison, 1415 University Ave, Madison, WI 53706
[3]Department of Materials Science and Engineering, University of Wisconsin – Madison, 1509 University Ave, Madison, WI 53706
[4]Department of Electrical & Computer Engineering, University of Wisconsin – Madison, 1415 Engineering Dr., Madison, WI 53706

ABSTRACT

Graphene plasmonic resonators have been broadly studied in the terahertz and mid-infrared ranges because of their electrical tunability and large confinement factors which can enable dramatic enhancement of light-matter coupling. In this work, we demonstrate that the characteristic scaling laws of graphene plasmons change for smaller (< 40 nm) plasmonic wavelengths, expanding the operational frequencies of graphene plasmonic resonators into the near-infrared (NIR) and modifying their optical confinement properties. We utilize a novel bottom-up block copolymer lithography method that substantially improves upon top-down methods to create resonators as narrow as 12 nm over centimeter-scale areas. Measurements of these structures reveal that their plasmonic resonances are strongly influenced by non-local and quantum effects, which push their resonant frequency into the NIR (2.2 μm), almost double the frequency of previous experimental works. The confinement factors of these resonators, meanwhile, reach 137 $\pm$ 25, amongst the large reported in literature for an optical cavity. While our findings indicate that the enhancement of some 'forbidden' transitions are an order of magnitude weaker than predicted, the combined NIR response and large confinement of these structures make them an attractive platform to explore ultra-strongly enhanced spontaneous emission.

INTRODUCTION

Optoelectronic devices based on tunable graphene plasmonic resonances offer a promising route towards the development of next generation devices[1-3] that include chemical sensors,[4-6] perfect absorbers,[7,8] tunable filters,[9] and high-speed intensity and phase modulators.[10,11] These devices operate by patterning graphene sheets into nanoribbons,[5-7,10,12,13] nanoperforated sheets,[4,14,15] or nanodisks[9,14] that support plasmonic oscillations with a frequency set by the characteristic length of the nanostructure and carrier density of the graphene.[12,14,15] Thus far, these devices operated at mid- and far-infrared wavelengths, and there have been both theoretical[16] and experimental[17,18] investigations that have indicated that the plasmonic response of graphene should be heavily damped at free space wavelengths shorter than 6 μm, where plasmon-phonon interactions can drive loss. Nevertheless, graphene plasmonic resonances at wavelengths as short as 3.5 μm have been observed,[12] and it is not yet clear if there is a high frequency cut-off for graphene plasmons and, if so, what sets that limit. If the plasmonic response of graphene could be pushed to shorter wavelengths and into the near-infrared (NIR), there would be interesting fundamental and technological implications. For example, theoretical models predict and experiments demonstrate that graphene plasmons could create extremely high Purcell enhancement rates of emission,[12,19] and could even be used to drive optically forbidden trasitions.[20] These processes are expected to become more dramatic in the NIR and, unlike the mid-infrared, there are numerous fluorescent NIR emitters – including quantum dots,[21,22] lanthanide ions,[23] and III-V materials[24] – that can be used as a platform to observe these effects. Meanwhile, from a technological perspective, a tunable NIR plasmonic response would allow for the creation of high-speed, low cost devices that operate at telecommunication wavelengths.

To-date, however, no experiments have demonstrated a graphene plasmonic response in the NIR, largely due to graphene's unique plasmonic dispersion relation, which leads to a massive mismatch between the plasmonic ($\lambda_p$) and free-space wavelengths ($\lambda_0$).[3] For the mid-infrared (MIR), this mismatch – or confinement factor ($\lambda_0/\lambda_p$) – is typically around 50 – 100, such that to achieve a plasmonic resonance at $\lambda_0$ = 5 µm, nanostructures at 30 nm length scales are needed (the nanostructure width approximately defines $\lambda_p/2$).[12] In the NIR, however, the confinement factor has been predicted to increase to 100-200, [16] and a device operating at $\lambda_0$ = 2 µm would require patterning the graphene down to 8 nm. Creating periodic patterns with these length scales over large (>100 µm²) areas is difficult for top-down lithography methods such as electron-beam lithography (EBL) - which was previously used to create plasmonic resonators down to 15 nm[12] - due to proximity effect distortions. Thus, to explore the NIR regime, new lithographic methods must be developed capable of creating graphene nanostructures at sub-15 nm length scales over centimeter-scale areas. A promising alternative to EBL is block copolymer (BCP) lithography, a bottom-up method that creates etch masks with dense nanoscale characteristic features over wafer-scale areas. BCP lithography has been explored as a means of patterning graphene into nanostructures including graphene nanoribbons (GNR) [25,26] and nanoperforated graphene (NPG).[4,27,28]

In this paper, we show that graphene nanostructures fabricated using BCP lithography on centimeter-scale areas of graphene can act as resonant plasmonic cavities. Using this method, we are able to produce graphene nanostructures with 12 nm length scales, that exhibit tunable plasmonic resonances up to 2.2 µm - the shortest wavelength yet reported in literature with a previous record of 3.5 µm.[12] Additionally, the confinement factors of our devices reach up to 137, matching the highest experimentally demonstrated value of an optical cavity in a 2D material.[12,29] The general width and carrier density-dependencies of these localized resonances are characteristic of graphene plasmonic devices that operate at longer wavelengths, namely that reduced width and increased carrier density both blue-shift the plasmonic resonance. However, we observed a larger than expected blue-shift in the measured resonant frequency of our smallest resonators in comparison to calculations given by a local electrodynamic model. We attribute these differences to both non-local and electron quantization effects in the nanoribbons.[30-32] These findings show that the characteristic scaling laws of graphene plasmons shift in smaller structures, leading to smaller confinement factors than predicted while simultaneously enabling the graphene plasmonic devices to operate at NIR frequencies.

RESULTS

**Graphene Nanostructure Fabrication**

We fabricated resonant graphene plasmonic devices using a BCP lithographic process outlined in Figure 1a, where a self-assembled polymeric material acts as an etch mask for pattern transfer. In general, the shape and size of the self-assembled polymer microdomains can be controlled by the molecular weights and relative volume fractions of each block.[33,34] We selected poly(styrene-*b*-methyl methacrylate) (PS-*b*-PMMA) to fabricate graphene nanostructures on $SiO_2$ substrates due to the nearly equal surface energies of PS and PMMA blocks, which makes thin-film assembly feasible, and the good etch selectivity of PMMA to PS in reactive ion etching (RIE) procedures used for pattern transfer to graphene. We use a substrate-independent neutral layer chemistry developed by Han *et.al.*[35] based on a random copolymer (RCP) poly(styrene-*r*-methyl methacrylate-*r*-glycidyl methyl methacrylate), abbreviated P(S-MMA-GMA). This RCP is spin coated onto graphene treated with pyrenebutyric acid (PBA),[36] which acts as an adhesion promotor (see SI section 1). Finally, the BCP dissolved in toluene is spin coated over the RCP-modified graphene and thermally annealed under vacuum to induce perpendicular BCP self-assembly.

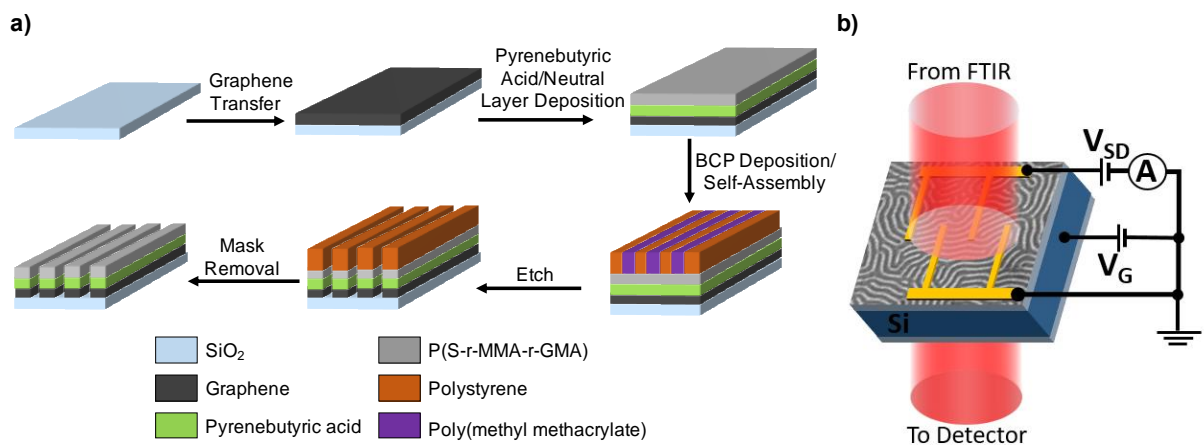

Figure 1: (a) Overview of the block copolymer procedure for fabricating graphene nanoribbons (GNRs). Procedure for nanoperforated graphene (NPGs) is identical, except we selected a BCP with weight fractions of the two blocks to induce cylindrical phase self-assembly. (b) Schematic of experimental device with electronic configuration shown for back-gated sample and a GNR SEM image embedded in the schematic.

We transferred the nanoscale BCP features to the graphene via an oxygen plasma RIE process, where the PMMA etches twice as fast as PS.[37] We attempted to remove the BCP mask using either organic solvents, thermal methods, or longer plasma etching, however, due to the relatively poor adhesion of graphene to $SiO_2$,[38,39] we were unable to remove the polymeric mask without losing adhesion of these graphene nanostructures to $SiO_2$ (see SI section 3). An optimized oxygen plasma etch time brought the residual polymer on the graphene nanostructures down to a couple of nanometers as confirmed with atomic force microscopy (AFM) (see SI section 2).

We fabricated two types of graphene nanostructures: NPG as regular hole arrays (Figure 2a) and GNRs in a 'zen-garden' pattern (Figure 2b). The characteristic widths of our final structures range from 29 ± 3 nm down to 12 ± 2 nm, as measured using both scanning electron microscopy (SEM) and atomic force microscopy (AFM) (See SI section 2). The minimum ribbon width of 12 nm represents the thermodynamic limit of self-assembly for PS-*b*-PMMA, but smaller features are theoretically achievable using more complex BCP compositions.[33,37] We also highlight that there is no significant difference in processing between a small or a large area of graphene as the spin coating and thermal annealing of the BCP employed here is applicable to wafer-scale substrates.

**Feature Size and Carrier Density Dependence of Graphene Plasmonic Resonators**

A cartoon schematic with an embedded SEM image of the GNRs and optical image of a GNR device with gold electrical leads is shown in Figure 1b. As-prepared CVD graphene sheets were hole-doped due to the iron chloride transfer process, with the background Fermi level ($E_F$) ranging from 0.4 - 0.5 eV. To tune the doping, we used either an electrostatic back-gate (see Figure 1b for the electrical setup), which allowed measurements up to $E_F$ = 0.59 eV, or a liquid ionic gel (see SI section 4), which allowed measurements up to $E_F$ = 0.74 eV.[40] The Fermi level in these experiments was determined by measuring the interband-transition spectrum of the sample, which displays a characteristic decrease in transmission at 2*$E_F$ (see SI section 5).[41] Plasmon resonances in these samples were probed by measuring their differential transmission spectrum using a Bruker Vertex 70 FTIR with Hyperion 2000 microscope. Doping-dependent spectra on each sample were normalized to spectra obtained at a Fermi level of 0 eV, the charge neutral point, taken from the same area of the sample. For some NPG samples, only the background doping was used, and no electrostatic gate was applied; for those samples, the spectra were normalized against unpatterned graphene (see SI section 7).

Figure 2 shows the width-dependent spectra of plasmonic resonances in both NPGs and GNRs created using BCP lithography, for a constant Fermi level of ~0.5 eV achieved via background doping for the NPGs and electrostatic doping for the GNRs. The resonances appear as dips in the normalized transmission spectrum, plotted as 1- (*normalized transmission*) to form peaks. The wavelength of these resonances show a distinct blue-shift as the width of both NPG and GNR dimensions are systematically lowered, which is consistent with previously reported plasmonic resonance in graphene nanostructures.[6,12,15] For these measurements, we define the widths of the NPG by the smallest distance between two holes (see Figure 2a) and the width of the GNRs is determined by the ribbon width (see Figure 2b). Variance in the width, as determined by SEM images, is measured to be approximately 2 to 3 nanometers, which creates a large fractional change in the width of the smallest features. This effect leads to substantial inhomogeneous broadening in plasmonic resonances for smaller GNRs and NPGs. Other sources of inhomogeneous broadening may include regions with poor electronic contact and uneven background doping.

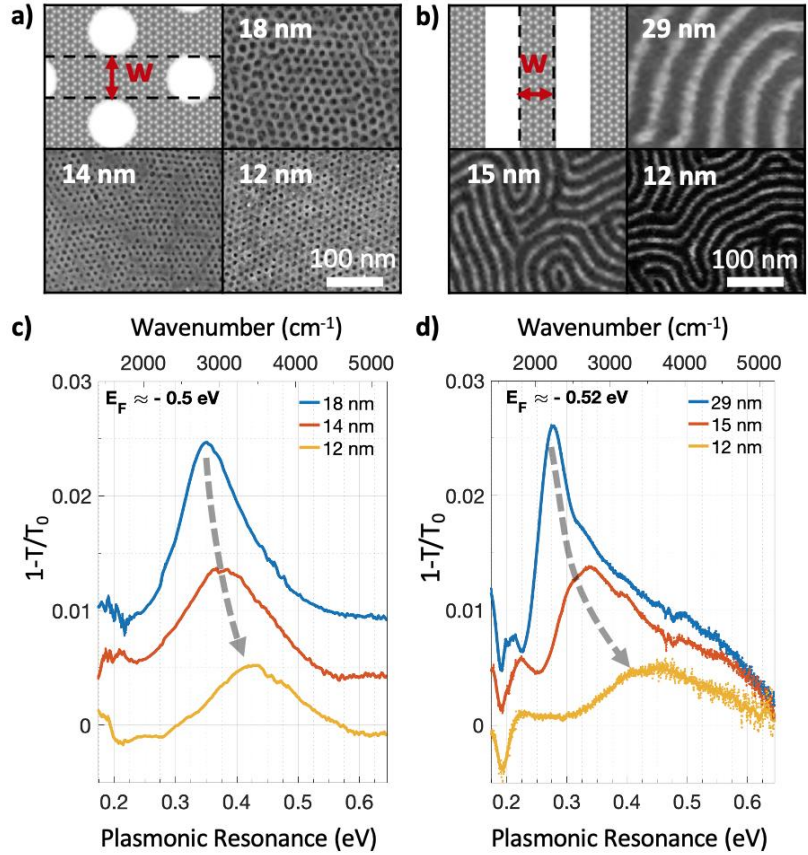

Figure 2: Modulation of transmission through graphene nanostructures by varying the widths at similar doping levels. (a) Schematic of nanoperforated graphene (NPGs) and SEM images containing 18 ± 3 nm, 14 ± 2 nm, and 12 ± 2 nm widths. (b) Schematic of graphene nanoribbons (GNRs) and SEM images containing 29 ± 3 nm, 15 ± 2 nm, and 12 ± 2 nm widths. (c) Modulation of transmittance through chemically doped NPGs normalized to transmission through unpatterned graphene. (d) Modulation of transmission through gate-induced doping of GNRs normalized to transmission at the charge neutral point. Gray arrows follow the peak shift as the width is changed.

Next, we measured the carrier-density dependence of these samples, as shown in Figure 3. These measurements show that increased doping leads to a blue-shift in the plasmonic resonant frequency, and an increase in the intensity of the spectral features for both NPG (Figure 3a) and GNR (Figure 3b, 3c) geometries. In Figure 3b, we tuned the plasmonic resonance to frequencies as high as 3800 cm$^{-1}$ (2.6 µm) with an electrostatic gate in a sample with 12 nm GNRs. In Figure 3c, our GNRs were of similar widths (13 nm vs 12 nm), but we were able to achieve larger doping levels with the ionic gel and consequently were able to tune the plasmonic resonance to approximately 4500 cm$^{-1}$ (2.2 μm). These results represent the first direct measurement of graphene plasmons at NIR frequencies. We note that the plasmonic resonance shift as a function of doping level is significantly smaller in the backgated samples than the ionic gel. We attribute these effects to inhomogeneous doping in the samples, which changes the local Fermi level. This is a problem in the backgated samples as we observe an averaging effect of multiple Fermi levels in measured resonances of the backgated samples, whereas the ionic gel can screen these charge inhomogeneities and maintain a constant Fermi level across the sample (see SI section 6).

We can further analyze the properties of these plasmonic modes by calculating their confinement factors, or the ratio between the free space wavelength and the plasmonic wavelength, $\lambda_0/\lambda_p$. We extracted the plasmonic wavelength by modeling the nanostructures as Fabry-Perot oscillators with a non-trivial edge scattering phase.[12] The resulting confinement factors are shown in Figure 4a for different ribbon widths and doping levels; we found that amongst our smallest ribbons, 13-nm wide GNRs, had confinement factors reaching approximately $137 \pm 25$. This value is 25% higher than what has previously been achieved in 20-nm wide GNRs,[12] and within a margin of error of the largest observed confinement factor in an optical cavity of a 2D material, which was demonstrated using vertically-confined acoustic graphene plasmons.[29]

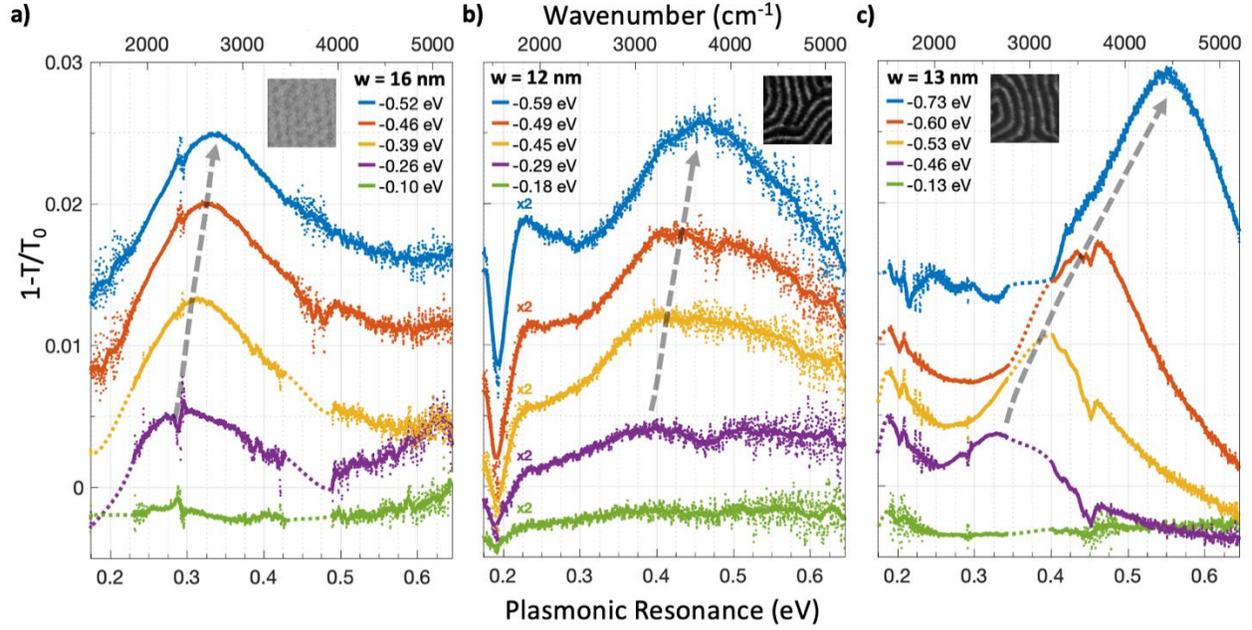

Figure 3. Modulation of transmission through graphene nanostructures by electrostatically doping the graphene. All measurements normalized to transmission at charge neutral point. (a) Backgate-induced doping of 16 ± 3 nm wide NPGs. (b) Backgate-induced doping of 12 ± 2 nm wide GNRs. (c) Ionic gel top gate-induced doping of 13 ± 2 nm wide GNRs. SEM insets show the structure measured in a, b, and c. Gray arrows follow the peak shift as the Fermi level is shifted.

The plasmonic resonances we observe display doping- and width-dependent frequencies that are qualitatively similar to theoretical predictions and experimental measurements of larger graphene nanostructures.[12,19] These behaviors can be observed in Figure 4b, where we plot the resonant frequencies of three representative GNR devices as a function of width for different Fermi levels. For comparison, we also plot the theoretical width-dependent resonant frequencies simulated using a first-order RPA model for graphene's conductivity with a finite-difference time-domain solver (solid lines).[42] We note that as the size of the resonator was reduced, the tunability of the plasmon dramatically increased. These results show that for our measurements widest resonators (purple triangles) there is good experimental and theoretical agreement, namely that the resonances occur at the expected Fermi levels given our uncertainties. However, our measurements reveal that when the GNR width is decreased below 15 nm, the plasmonic resonances occur at energies that diverge from the behavior observed in larger resonators and are blue-shifted from calculations based on the first-order random phase approximation (RPA).[42] To illustrate, consider the 13-nm GNRs at $E_F$ equal to 0.60 eV in Figure 4b. Instead of intersecting the 0.6 eV $E_F$ curve, it lies on the 0.8 eV $E_F$ curve. Furthermore, the plasmonic resonance at $E_F$ equal to 0.71 eV for the same sample is predicted by the local model to occur at roughly $\omega_p$ equal to 0.45 eV instead of the measured resonance at 0.55 eV, an almost 20% increase. In order to match the theoretical curves of the local model, the widths of the GNRs would need to be nearly half the widths measured using SEM (see SI section 2), or the carrier density would need to be at least 30% more than what we estimate based on the experimentally measured interband transition cutoff (see SI section 5).

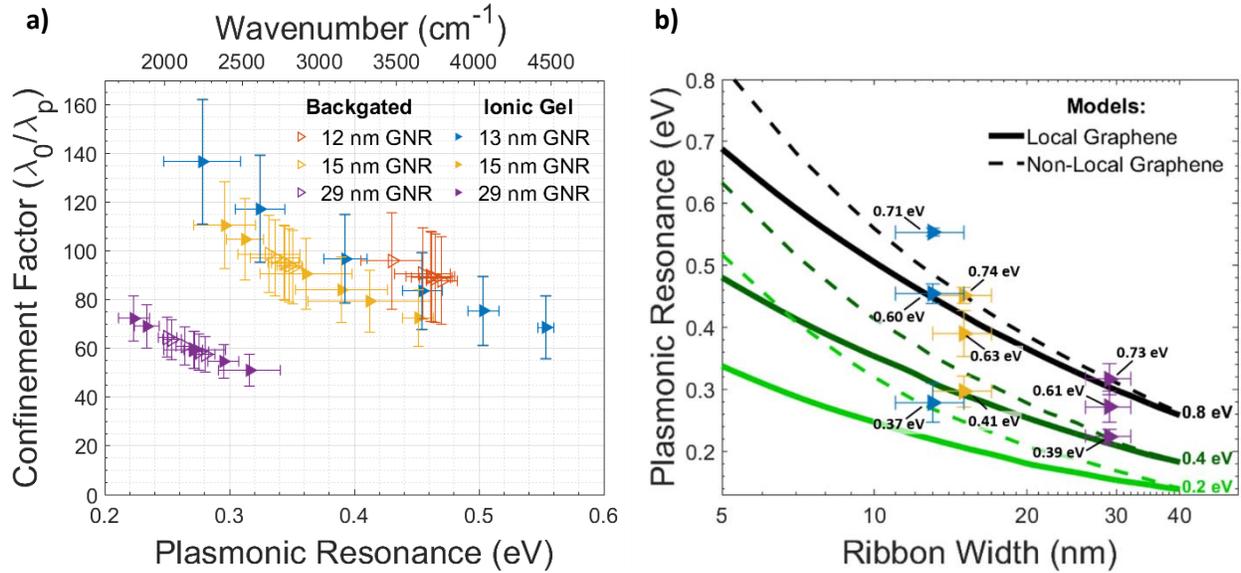

Figure 4: (a) Confinement of graphene plasmons in different widths GNRs plotted as a function of plasmon resonance. (b) Theoretical dispersion of GNR plasmons as a function of ribbon width modeled using the classic local approximation (solid) and including a non-local shift (dashed), for various Fermi levels with the Fermi levels of the measurements denoted on the plot. Open and filled symbols plot the measured energy of the plasmons of the GNRs doped by a backgate or ionic gel, respectively. Symbol color corresponds to specific samples and is consistent between both plots.

DISCUSSION

We conclude that the observed offsets are not due to potential experimental errors, but rather can be understood as non-local and electron quantization effects perturbing the graphene plasmons, both of which occur more dramatically as the GNR width is decreased.[30-32,43] The first one of these effect, optical non-locality, describes the effect of an electric field in one location producing a polarization in the nearby vicinity, such as through the Coulomb interactions of charge carriers. Such perturbations are known to occur most strongly in regions of high optical field enhancement, as in metallic nanostructures with small (< 10 nm) geometric features.[44,45] We incorporate non-locality into our theory with a hydrodynamic term in the RPA description of the graphene dielectric response, and use a finite-element Maxwell equation solver (COMSOL) to calculate the resonant frequency shifts for different ribbons widths and doping levels. This has previously been derived as a modification to Ohm's Law, which connects the in-plane currents $\boldsymbol{J}$ and electric fields $\boldsymbol{E}$ as follows: $\boldsymbol{J} = \sigma(\omega)\boldsymbol{E} - \frac{\beta^2}{\omega^2}\nabla(\nabla \cdot \boldsymbol{J})$, $\sigma(\omega)$ is the conductivity of graphene and the $\beta^2$ term accounts for the pressure of an inhomogeneous electron fluid, representing the approximation of the non-local effects.[30,45] The results of those calculations (shown in Figure 4b as dashed lines) indicate that non-local effects blue-shift the resonant plasmon energies, and that this effect is more dramatic as the ribbon width is decreased, and for lower doping levels. For example, for 29-nm GNRs at an $E_F$ equal to 0.73 eV, non-local effects shift the resonant frequency by 2%, while for 13-nm-wide GNRs at an $E_F$ equal to 0.71 eV, the resonant frequency is shifted by 12%. While these corrections bring the theoretical predictions closer to our experimental observations, they still do not fully explain the deviations we observe. The second possible contribution we consider is electron-quantization in the GNRs, where the continuum model for graphene quasiparticles breaks down, forming low energy band gaps and exotic edge states that depend on the exact edge termination of the GNRs.[46-51] Such effects are predicted to blue-shift the plasmonic resonance by up to 5% for 13 nm wide GNRs and thus may also contribute to the overall shift we observed.[31,32] The prospect of electron quantization occurring in these devices is supported by transport measurements performed on similarly prepared NPG samples with characteristic widths of 18 ± 2 nm.[27] Those measurements revealed an effective electronic bandgap of 100 meV, which could strongly perturb the infrared graphene dielectric properties, and drive changes in the plasmonic response. While quasiparticle quantization and optical non-locality could both contribute to the phenomena we observe in this study, a complete theoretical model that includes both effects is beyond the scope of this work and left for future investigation.

      The blue-shift in resonant frequency that we observe has broad implications for the limits of future optoelectronic devices based on graphene plasmons. Most importantly, our results reveal that

graphene nanostructures can exhibit a strong, tunable optical response in the NIR. Moreover, the length scales needed to realize such behavior are larger than previously predicted, and directly accessible to block copolymer lithography. For example, our first-order RPA calculations indicated that in order to show plasmonic activity at telecom frequencies (1550 nm), GNR widths of 4.5 nm at a carrier density of 1 eV would be necessary. Because the non-local and quantum contributions increase in magnitude as the width is decreased, it may be possible to reach telecom frequencies in GNRs with 7-10 nm widths – a more reasonable, but still a challenging fabrication-limited length scale to reach. Another important consequence of the blue-shift that we observe is that the confinement factors of graphene plasmons at short wavelengths ($\lambda < 4$ μm) are much smaller than previously predicted.[16] This decrease in confinement indicates that several predicted phenomena that leverage graphene plasmons – including SPASing,[52] enhanced spontaneous emission of forbidden transitions,[20] and enhanced sensing[5] – may occur at lower rates than previously anticipated. As an example, for 13 nm GNRs, a confinement factor of ~120 at $E_F$ equal to 0.64 eV was expected from previous calculations,[16] in comparison to the 84 $\pm$ 13 confinement factor we observed at $E_F$ equal to 0.63 eV. Since two-plasmon spontaneous emission rates scale as $\left(\frac{\lambda_0}{\lambda_p}\right)^6$,[20] our results indicate emitters coupled to graphene will undergo two-plasmon spontaneous emission at a rate that is one order of magnitude lower than previously thought.

In conclusion, we developed a fabrication technique based on block copolymer lithography that allowed us to probe new regimes of graphene plasmonics with the first measurement of graphene plasmons in the NIR. Our results provide direct evidence that scaling laws of graphene plasmons change due to both non-local and electron quantization effects. Consequently, reaching the NIR and shorter wavelengths can be achieved in larger structures than previously thought, enabling a new avenue of research for graphene-based NIR devices. We fabricated graphene nanostructures that, to the best of our knowledge, exhibit among the largest lateral confinement factors as optical cavities in the literature.[29,53-56] However, these results demonstrate confinement will be less than previous predictions,[16] limiting the anticipated ability of graphene to enhance light-matter interactions. We note that in contrast to top-down methods commonly employed, our technique is designed to fabricate centimeter-scale areas of graphene with nanometer-scale precision, allowing for unprecedented ease of integration of these nanostructures into future optoelectrical devices. Graphene plasmonic devices created using BCP lithography represent an exciting platform to expand the tunable working range of graphene plasmonics, enhance light–matter interactions, explore quantum effects, and create new types of graphene devices.

METHODS

***Graphene CVD Growth.*** Graphene was grown by chemical vapor deposition methods on low purity Cu foil (15465015, Alfa Aesar). Cu foil was pre-cleaned with a 30 s 3:1 deionized water:ammonium persulfate dip (APS – 100, Transene). Graphene was grown at 1035 °C for 90 minutes while 42 sccm $H_2$ and 0.2 sccm $CH_4$ (both Air Gas Research 6.0 Grade) flowed in the quartz growth tube (260 mTorr). To transfer the graphene after growth, graphene was first coated with a protective layer of PMMA (950k A4, MicroChem Corp.) and the Cu foil was etched away with $FeCl_3$ (CE – 100, Transene). The graphene/PMMA stack was then rinsed in a series of deionized water baths. $Si/SiO_2$ substrates (double side polished, 285 nm dry thermal oxide, prime grade Si wafers from WaferPro) were piranha-treated using 3:1 $H_2SO_4$:$H_2O_2$ at 85 °C for 30 min immediately prior to graphene/PMMA transfer. Once transferred, the PMMA was removed by soaking in 60 °C acetone for 1 hr and the graphene/$SiO_2$ was annealed for 1 hr at 400 °C under $10^{-6}$ Torr vacuum.

***Nanoribbon and Nanoperforated Graphene Fabrication.*** The monolayer graphene coated $SiO_2$ substrates were placed in a 1-pyrenebutyric acid (PBA) (257354, Sigma Aldrich) solution in THF for 24 h based on a previously developed procedure.[36] A random copolymer (RCP) of glycidyl methacrylate (GMA), styrene (S), and methyl methacrylate (MMA), P(S-*r*-MMA-*r*-GMA), was synthesized as reported earlier.[57] For this study, the RCP contained 62.5% S, 4% GMA, and balance PMMA (as confirmed by NMR) for lamellar forming BCPs. For cylindrical BCPs, the RCP contained 72% S, 4% GMA, and balance PMMA. The RCP was dissolved in toluene (320552, Sigma Aldrich) and spin coated on the PBA-coated graphene/$SiO_2$ substrate. These samples were annealed at 160 °C for 3 hr under vacuum to crosslink the GMA unit and soaked in toluene for 15 min to remove any unreacted RCP. We used BCP P(S-*b*-MMA) with various molecular weights (MWs) all from Polymer Source, Inc., specifically 17k-17k and 53k-54k

MWs for lamellar-forming PS-*b*-PMMA and 21.5k-10k MW for cylindrical-forming PS-*b*-PMMA. These BCPs were prepared in toluene and spin coated onto the random copolymer covered samples. Films were thermally annealed under vacuum for BCP self-assembly. Pattern transfer from the BCP to the underneath graphene was performed using a reactive ion etcher (Plasma-Therm 790) with oxygen plasma. Scanning electron microscope (SEM) images were taken with a Zeiss LEO 1550VP SEM for nanopatterned graphene visualization. (Back-gated only NPG samples were fabricated with a slightly modified fabrication procedure described in detail in SI Section 7. Additional samples were prepared using these procedures that exhibited optical behavior similar to what was reported in the main text.)

***Fabrication of Gold Contacts.*** For the GNR coated $SiO_2$/Si substrates, they were first coated with a bilayer PMMA (950 A4 at 200 nm and 495 C2 at 300 nm). These samples were then exposed and patterned using the Elionix ELS G-100, an electron-beam lithography tool. After exposure, the substrates were developed in a 3:1 IPA:DI water mixture for 20 s with a 10 s rinse of IPA. Metal deposition consisted of a 2.5 nm chromium adhesion layer and 80 nm of gold. After deposition, the PMMA was removed via a 15 hr chloroform bath, a 1 hr acetone bath, then a 5 min IPA bath with a 5 s IPA rinse. The contacts for the GNRs were a gold-electrode mesh of interlocking branches that are 80 nm apart and spaced by 3 μm. This pattern ensures reliable contact to the graphene nanoribbons in the `zen-garden' pattern without significantly reducing the optical signal or introducing additional resonances in the wavelength range of interest (2-6 μm). For NPGs, the graphene remained a continuous sheet, so large gold contacts were applied via metal deposition through a mask.

***Ionic Gel Preparation***

Ionic gel preparation based on previously established procedure in literature.[40] Diethylmethyl(2-methoxyethyl)ammonium bis(trifluoromethylsulfonyl)imide ([DEME][TFSI]), was purchased from Sigma-Aldrich Chemicals. Initially, the ionic liquid was dried at 105 °C under vacuum for 3 days. Polystyrene-b-poly(ethylene oxide)-b-polystyrene (PS-*b*-PEO-*b*-PS) triblock copolymer was purchased from Polymer Source, Inc. The molecular weight of the block copolymer was PS 10 kg/mol, PEO 44 kg/mol, and PS 10 kg/mol which corresponds to PEO volume fraction of 0.67. 0.55g of [EMIM][TFSI] was dissolved with 21 mg of PS-*b*-PEO-*b*-PS in 2 mL of anhydrous dichloromethane. The solution was stirred overnight at room temperature. The ionic gel was spin-coated on the graphene sample and stored under $N_2$ until ready for use.

***FTIR Measurements.*** The transmission measurements were performed using a Bruker Vertex 70 FTIR attached to a Hyperion 2000 microscope in a nitrogen-purged environment. We used a liquid-nitrogen-cooled mercury–cadmium–telluride (MCT) detector. For our plasmonic resonance measurements, we used a potassium-bromide (KBr) beam splitter with a silicon carbide glow bar as our MIR source. For the measurements of the Drude peaks (see SI section S5), we used a quartz beam splitter with a halogen bulb as our source.

***Electromagnetic Simulations.*** We rigorously solved Maxwell's equations using both the finite-difference time-domain (FDTD) method and the finite-element method (FEM). While these methods give the same result for the local calculations, the FDTD simulations were used as they ran faster than the FEM calculations in our set up. However, our FDTD solver could not incorporate non-local corrections, so we switched to an FEM solver for the non-local calculations. Graphene was modeled as a thin material of thickness $\delta$ with permittivity $\epsilon_G = \epsilon_r + \frac{i\sigma(\omega)}{\epsilon_0 \omega \delta}$. $\sigma(\omega)$ is the complex optical conductivity of graphene evaluated with the local random phase approximation.[42] $\delta$ was chosen as 0.3 nm which showed good convergence with respect to the $\delta \to 0$ limit and $\epsilon_r$ is the background relative permittivity. For the FDTD simulations, we used the commercial solver Lumerical FDTD. The graphene nanostructures were simulated on an SiO$_2$ layer with thickness of 285 nm, on top of a Si substrate. Material properties from Palik[58] were used for both materials. To model the structure accurately, we included a polymer layer of thickness 1.2 nm, with dielectric properties taken as average of PMMA[59] and PS.[60] Our simulations of the non-local effect made use of the FEM solver COMSOL Multiphysics. We made use of the NanoPL RF module,[45] an extension designed for calculating non-local effects of 2D nanostructures. Graphene was modeled identically to the FDTD approach and an average background dielectric of 2.5 was used to represent the contributions of the dielectrics surrounding the graphene to estimate the shift caused by the non-local effects.

**Author Contributions**

J.F.S. and J.H.D. contributed equally to this project. M.F., N.S.S., and V.S. grew graphene. J.H.D., A.S., J.W.C., and W.W. fabricated graphene nanostructures with block copolymer lithography procedures. J.H.D., J.W.C., and N.S.S. performed SEM imaging. J.H.D. and A.S. characterized graphene nanostructure size with SEM images. W.A.B. performed AFM imaging and graphene nanostructure size analysis with AFM images. J.F.S. fabricated electrical leads on graphene nanostructures. J.F.S. and C.W. performed FTIR measurements and analysis. J.F.S. performed electromagnetic simulations. J.F.S. and J.H.D. lead manuscript preparation, with contributions from all authors. V.W.B., P.G., M.A.K., and M.S.A. supervised this research project.

**Acknowledgments**

J.F.S., W.W., P.G., and V.W.B. were supported by a Defense Advanced Research Projects Agency Young Faculty Award (YFA D18AP00043). P.G. and J.H.D. received partial support from SNM-IS award No. 1727523. J.H.D. acknowledges support from PPG Industries, Inc. through a fellowship program at University of Wisconsin-Madison. N.S.S. and M.S.A. were supported by U.S. Army Research Office, W911NF-12-1-0025 and then W911NF-18-1-0149. V.S. and M.S.A. were supported by the U.S. Department of Energy, Office of Science, Basic Energy Sciences, under award No. DE-SC0016007. M.F. was supported by the Gordon and Betty Moore Foundation through a Moore Inventors Fellowship. W.A.B. is supported by the National Science Foundation under Grant No. DMR-1839199. C.W. and M.A.K. are supported by the Air Force Office of Research under Grant No. FA9550-18-1-0146. The authors gratefully acknowledge the use of facilities and instrumentation at the University of Wisconsin-Madison Wisconsin Centers for Nanoscale Technology partially supported by the NSF through the University of Wisconsin Materials Research Science and Engineering Center (DMR-1720415). Portions of this work were done using an electron beam lithography tool at the University of Wisconsin Nanoscale Fabrication Center. We acknowledge support from the NSF (DMR-1625348) for the acquisition of this tool.

**Supplementary Information to "Scalable near-infrared graphene plasmonic resonators exhibiting strong non-local and electron quantization effects"**

**S1. Pyrenebutyric Acid Interfacial Layer for Graphene and Polymers**

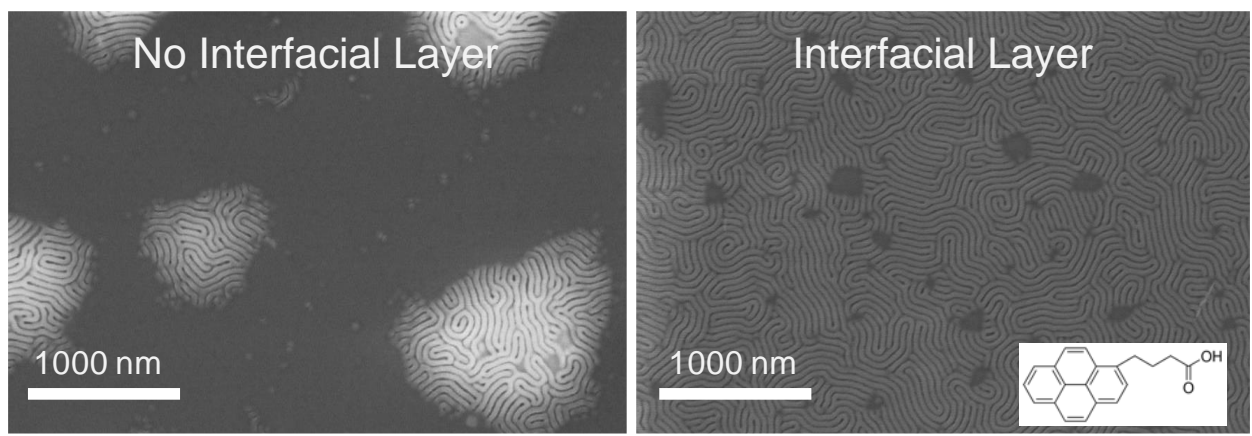

Figure S1: SEM images illustrating the effect on pyrenebutyric acid (PBA) on BCP self-assembly. Only the substrate with PBA in between graphene and the polymer thin film shows large area self-assembly.

## S2. Calculating Width of Graphene Nanoribbons

### S2.1. Calculating Width of Ribbons and Thickness via SEM

We measured the width of our graphene nanostructures using SEM image analysis. Using ImageJ software, we measured the width of these graphene nanoribbons by averaging 20 different ribbons. We report our errors as standard deviation plus pixel size image resolution added in quadrature. We considered ribbon width area using binary contrast transformation cutoffs in ImageJ. Using Fig. S2 as an example, we get a graphene nanoribbon width of 12 +/- 2 nm.

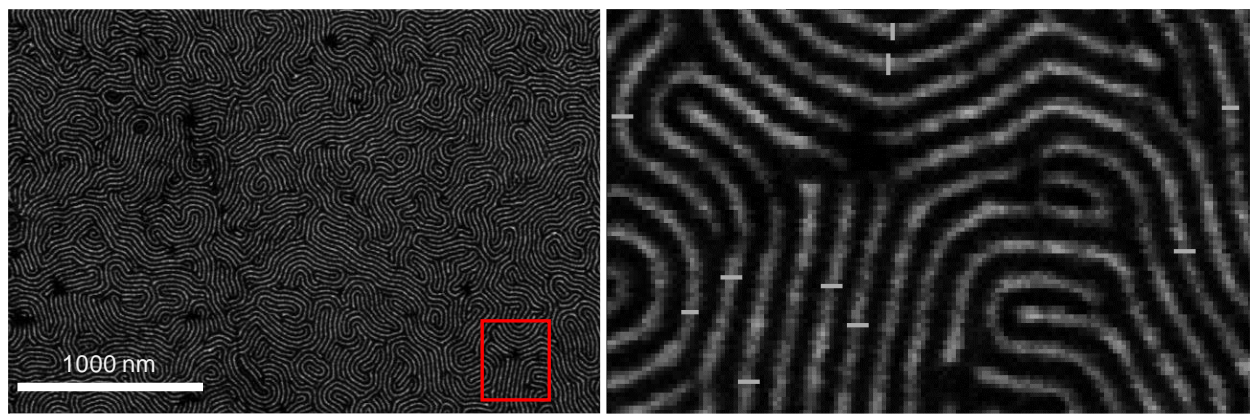

Figure S2: Representative SEM image of smallest graphene nanoribbons. We get an average ribbon width of 12 +/- 2, with our error representing standard deviation of averaged line widths plus pixel size image resolution added in quadrature.

### S2.2. Estimating Width of Ribbons and Thickness via AFM

We measured the height of our ribbons, including any residual mask polymer with an UHV cryogenic AFM at 4K. We can see that there is the finger-print pattern, indicative of the BCP ribbons. From this measurement extract the height difference between the valleys (no graphene) and the peaks (graphene with polymer) and an estimate of the ribbon thickness. Measurements were done using Gwyddion software. We found the height to be 2+/-0.5nm. Noting that graphene is approximately 0.3 nm thick, we estimate the height of the remaining polymer as 1.7 +/- 0.5nm. This means that the residue on the surface of the GNRs are the Pyrene and the Neutral layer. For this device, we found a ribbon width of 13.1 +/- 1.3 nm, comparable to the SEM measurements of the device at 13 +/- 2 nm.

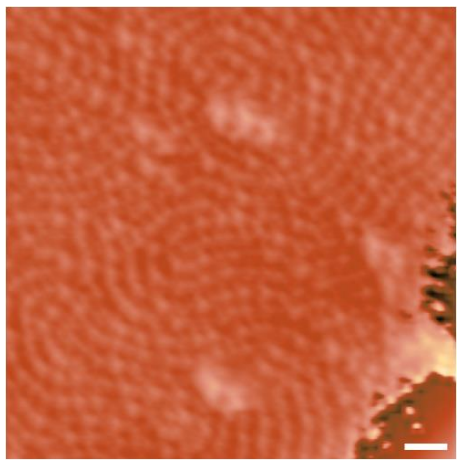

Figure S3: Representative AFM image of smallest graphene nanoribbons. Using Gwyddion software we measure the width of these graphene nanoribbons and the residual polymer height. Scale bare is 50 nm.

## S3. Removing Polymer Mask

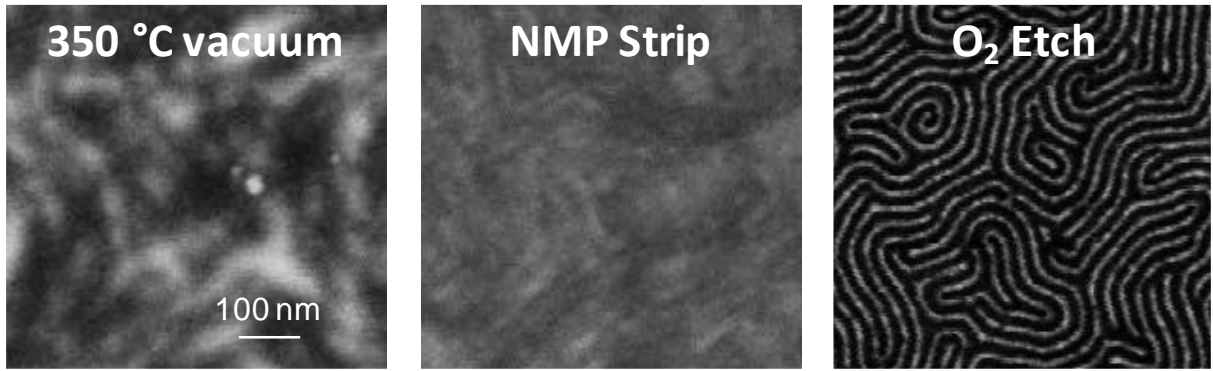

SEM images of various attempts to remove the polymer mask after pattern transfer to graphene. Scale bars are the same for all SEM images. Vacuum annealing left a large amount of residual polymer on the surface. Commercial photoresist strippers like AZ 400T NMP-based solvent removed more of the polymer mask, but still left residual polymer and regions of the substrate showed GNRs that delaminated from the $SiO_2$ surface. The oxygen plasma etch left a thin layer of polymer on the surface, but no graphene ribbons shifted during the removal process.

## S4. Ionic Gel

As an alternative to the backgate geometry discussed in the main text, we also made use of an ionic liquid to gate the graphene. In this method, the surface is coated with an ionic liquid (DEME) and the graphene is separated into two regions. These regions will act as our electrode and count-electrodes. By applying a voltage bias between the two electrodes, the ions in the ionic liquid will gather on one of the electrodes and charge of opposite sign will build up on the other set of electrodes, doping the graphene. We can measure the source/drain current across the electrode to measure the resistance of the graphene.

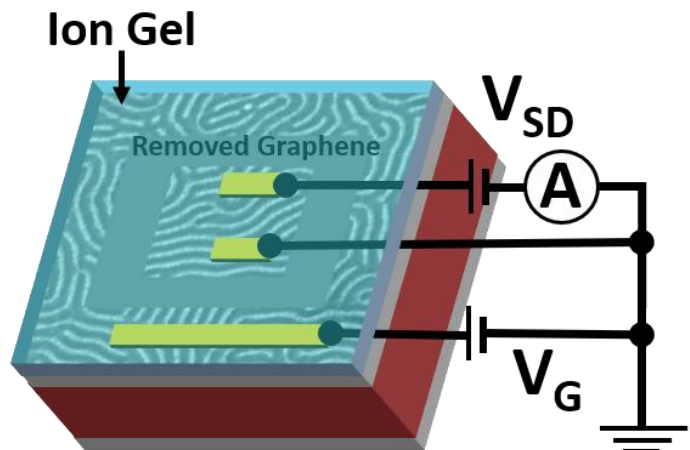

**Figure S4: Schematic of ionic gel top gate configuration.**

## S5 Estimating Fermi Levels

To understand the doping of our samples, we employed two approaches in tandem which find the charge neutral point and the Fermi levels directly.

### S5.1 Measuring the Charge Neutral Point

In the first approach, we varied the applied voltage (works for both backgated and ionic gel samples) until we found a resistance maximum, corresponding the charge neutral point . This is shown in Figure S5. One problem we experienced with backgating was that the CNP would be beyond the dielectric breakdown voltage of the oxide on sample, at approximately 200V - although the exact voltage was sample dependent. As the CNPs we were able to measure were around this breakdown voltage, we expect the CNPs of the other samples to be at similar values, but past the breakdown voltage of the sample.

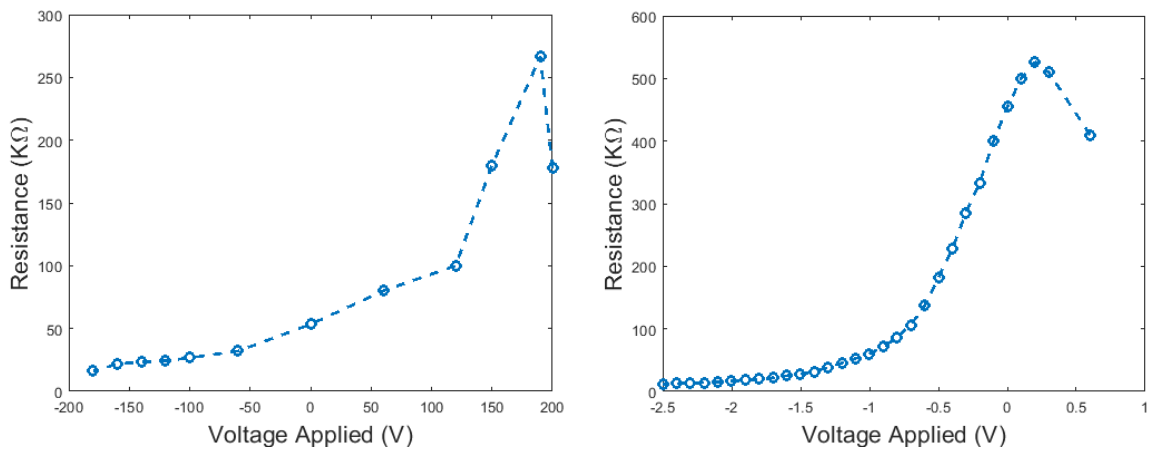

Figure S5: Representative resistance curves of a backgated (left) and ion-gel (right) doped samples. The presence of the peak in the curve corresponds to the CNP.

### S5.2 Measuring the Drude Peak

We can also directly measure the Fermi level of our graphene structures by measuring the Drude peak, seen in figure S6. Due to Pauli blocking, there will be increased transmission up to a threshold photon energy of $2E_F$, where interband transitions begin to occur. This feature can be identified as the region of maximum slope of the Drude peaks. Gating the graphene to increase the carrier density will shift location of this feature to a higher energy, as expected.

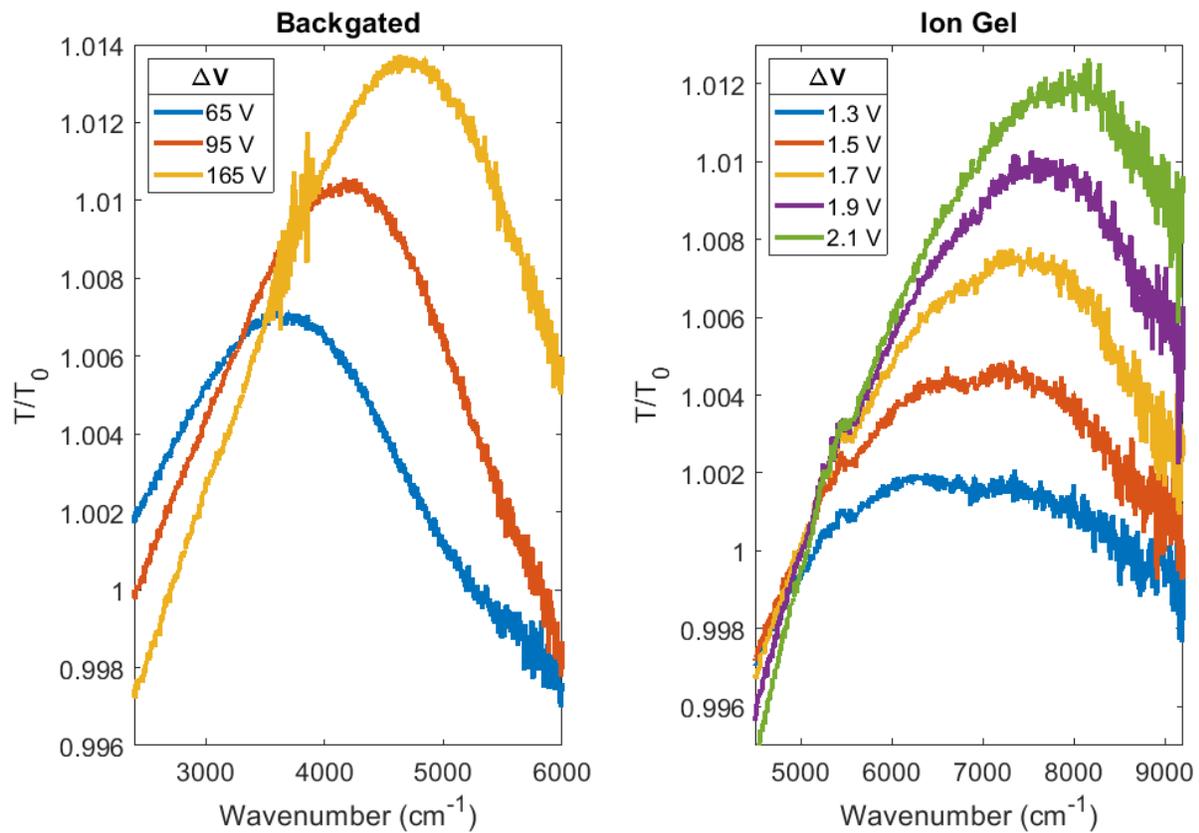

Figure S6: Representative Drude spectrum of a backgated (left) and an ionic gel (right) doped nanostructured graphene.

## S6. Modeling Reduced Plasmon Dispersion

As noted in the main text, we observed a crowding behavior where the estimated Fermi levels in backgated samples were much less dispersive than the theoretical Fermi levels based on the observed plasmonic resonances. We believe this effect to be due to the ribbons on a single sample not sharing the same background doping caused by our fabrication process introduction of charge inhomogeneity. This effect only occurred in our backgated samples as the ions in the ionic gel can screen charge inhomogeneities to maintain a constant CNP across the graphene eliminating the biasing effect. To model this effect, we simulated two sets of ribbons sizes (12 and 30 nm), each with three different background doping levels (0.4, 0.45, 0.5 eV), corresponding to CNP's at approximately 160, 200, and 245 V. For the gating measurements, we control the measurement voltage, but that will apply a different doping level dependent on the background doping of the graphene, which is detailed in Table S1.

| Measurement Voltage (V) | 120 | 0 | -120 |
|---|---|---|---|
| CNP (V) | | | |
| 245 | 0.356653 | 0.499314 | 0.609449 |
| 200 | 0.285322 | 0.451134 | 0.570645 |
| 160 | 0.201753 | 0.403507 | 0.533789 |

Table S1: Doping levels of the GNRs for different CNPs at different measurement voltages. The colors correspond to the lines in Figure S7.

In figure S7, we plot the weighted sum (our effective measured spectra) of the GNRs for three different measurement voltages. The left figure is for 12 nm ribbons, and the right is for 15 nm ribbons. For this representative example, we set the weighting of 0.4 eV to 2/3, 0.45 eV to 1/5, and 0.5 to 2/15. As we can see, our "measured" spectra have resonance peaks much closer together than the spectra of the ribbons at one Fermi level.

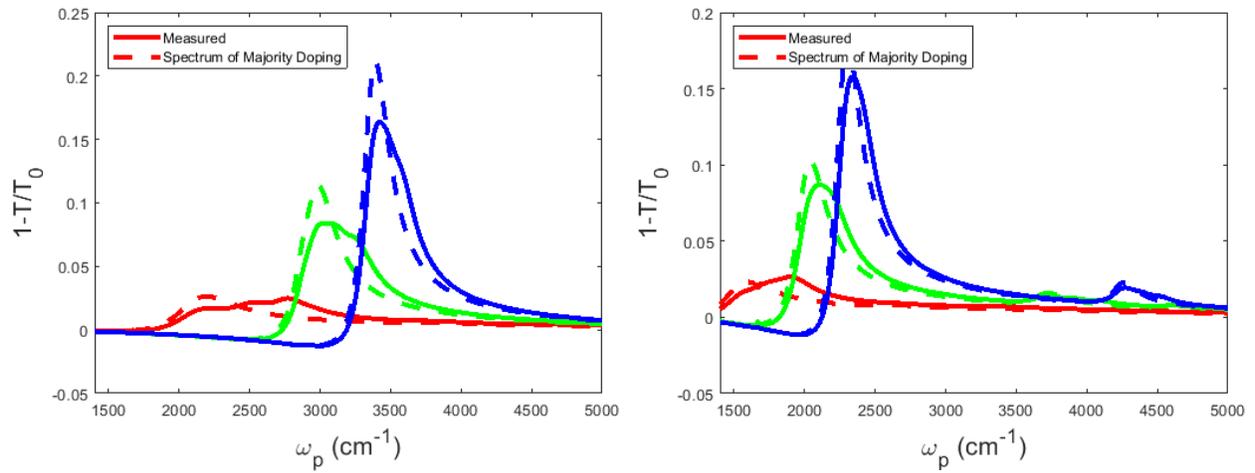

Figure S7: Comparison of spectra of GNRs with single background doping (solid), set to majority doping value, compared to spectra of the weighted sum (dashed lines). 12 nm (left) and 30 nm (right) were considered.

**S7. Additional Fabrication Details of Nanoperforated Graphene Samples.**

NPG samples without an electrostatic gate were fabricated using a slight variation of a procedure in the Methods published earlier.[1] Graphene was grown using thermal chemical vapor deposition of ultra high purity $CH_4$ at 1050 °C on Cu foil. As-received copper foils (Bean Town Chemical # 145780, 99.8% purity) were cut into 1 inch x 1 inch pieces and soaked in dilute nitric acid (5.7%) for 40 s followed by 3x DI water rinse followed by soaking in acetone and IPA to remove water from the surface. Dilute nitric acid helps remove the oxide and impurity-particles from the surface. Foils were then dried under a gentle stream of air. Foils were subsequently loaded into a horizontal quartz tube furnace in which the furnace can slide over the length of the tube. Prior to synthesis, the CVD chamber was evacuated to <$10^{-2}$ Torr using a scroll pump. The system was then back-filled with Ar and $H_2$, and a steady flow (331 sccm Ar, 9 sccm $H_2$) monitored by mass flow controllers was maintained at ambient pressure. The furnace was then slid to surround the samples, and annealed for 30 min. Then 0.3 sccm of P-5 gas (5% $CH_4$ in Ar) was flowed for 45 min so that a monolayer of graphene formed on the surface. To terminate the growth, the furnace was slid away from the samples, and the portion of the quartz tube containing the samples was cooled to room temperature.

PS-*b*-PMMA BCP films were first perpendicularly oriented on sacrificial 90 nm $SiO_2$/Si wafers from Addison Engineering, Inc. A random copolymer (RCP) of glycidyl methacrylate (GMA), styrene (S), and methyl methacrylate (MMA), P(S-r-MMA-r-GMA), was synthesized as reported earlier.[2] For this study, the RCP contained 70% S, 4% GMA, and balance PMMA as confirmed by NMR. The RCP was dissolved in toluene (320552, Sigma Aldrich) and spin coated on the PBA-coated graphene/SiO2 substrate. These samples were annealed at 160 °C for 3 hr under vacuum to crosslink the GMA unit and soaked in toluene for 15 min to remove any unreacted RCP. PS-*b*-PMMA BCP from Polymer Source, Inc. had molecular weight of 21.5k-10k. The BCP was dissolved in toluene and spin coated onto the random copolymer covered samples. Films were thermally annealed under vacuum for BCP self-assembly. The resulting BCP film was submerged in dilute (5%) hydrofluoric acid until the Si substrate detached from the BCP film resulting in a floating BCP film. This BCP film was transferred two times to DI water to remove trace amounts of HF and then transferred to a graphene coated 300 nm $SiO_2$/Si substrate. BCP films were etched with oxygen plasma to transfer BCP pattern to graphene.